\documentclass[Conference]{IEEEtran}

\usepackage{url}
\usepackage{graphicx}
\usepackage{subcaption}

\usepackage{amsfonts}
\usepackage{multicol}
\usepackage{booktabs}
\usepackage{textcomp}
\usepackage{comment}

\usepackage{mathtools}

\usepackage{stfloats}
\usepackage[below]{placeins}

\usepackage{cite}

\usepackage{pifont}

\usepackage{amssymb}

\usepackage{multirow}
\usepackage{amsmath}
\usepackage{csquotes}
\usepackage{color,soul}
\usepackage[bookmarks=false]{hyperref}
\usepackage{algorithm}
\usepackage{algorithmic}

\makeatletter
\def\ps@IEEEtitlepagestyle{%
  \def\@oddfoot{\mycopyrightnotice}%
  \def\@evenfoot{}%
}
\def\mycopyrightnotice{%
  {\footnotesize 978-1-7281-9226-0/20/\$31.00 \textcopyright 2020 IEEE\hfill}
  \gdef\mycopyrightnotice{}
}

\ifCLASSINFOpdf
\else
\fi

\begin{document}

\title{Matrix Decomposition for Massive MIMO Detection}

\author{\IEEEauthorblockN{Shahriar Shahabuddin\IEEEauthorrefmark{1},
Muhammad Hasibul Islam\IEEEauthorrefmark{1},
Mohammad Shahanewaz Shahabuddin\IEEEauthorrefmark{2},
Mahmoud~A.~Albreem\IEEEauthorrefmark{1}\IEEEauthorrefmark{3},
Markku Juntti\IEEEauthorrefmark{1}
\\}

\IEEEauthorblockA{\IEEEauthorrefmark{1}Centre for Wireless Communications, University of Oulu, Finland\\
\IEEEauthorrefmark{2}Vaasa University of Applied Science, Finland\\
\IEEEauthorrefmark{3}A'Sharqiyah University, Oman\\
}

\IEEEauthorblockA{ Email: \IEEEauthorrefmark{1}[firstname.lastname]@oulu.fi,   
\IEEEauthorrefmark{2}e1900302@vamk.fi,
\IEEEauthorrefmark{3}[firstname.lastname]@asu.edu.om}  
}

\maketitle

\begin{abstract}

Massive multiple-input multiple-output (MIMO) is a key technology for fifth generation (5G) communication system. MIMO symbol detection is one of the most computationally intensive tasks for a massive MIMO baseband receiver. In this paper, we analyze matrix decomposition algorithms for massive MIMO systems, which were traditionally used for small-scale MIMO detection due to their numerical stability and modular design. We present the computational complexity of linear detection mechanisms based on QR, Cholesky and LDL-decomposition algorithms for different massive MIMO configurations. We compare them with the state-of-art approximate inversion-based massive MIMO detection methods. The results provide important insights for system and very large-scale integration (VLSI) designers to select appropriate massive MIMO detection algorithms according to their requirement.
\end{abstract}


\IEEEpeerreviewmaketitle

\section{Introduction}
Massive MIMO is a key technology for fifth generation (5G) communication systems to achieve very high performance within the available radio spectrum. It is an extension of conventional small-scale MIMO, where a large number of antennas are used at the base station (BS) which serves a large number of users to achieve high data throughput and spectral efficiency~\cite{marzetta2010noncooperative,RPLLMET13}. Massive MIMO can provide uniform good service to the user terminals with a high mobility environment. However, the benefit of massive MIMO systems come with the disadvantage of computational complexity. The complexity of symbol detection algorithms grows exponentially as the number of antennas increase in a massive MIMO system~\cite{MIMO_survey}. Due to the high number of antennas, conventional linear detectors also require large matrix inversion. Hence, in the past decade, a new class of detectors based on approximate inversion has become a popular choice in very large-scale integration (VLSI) implementation. 

Approximate inversion-based detectors (AID) utilize channel hardening properties of massive MIMO systems to calculate the \textit{Gramian} matrix inversion. This class of detectors work well for certain configurations of massive MIMO systems, for example, when the ratio between numbers of BS antennas and users is large~\cite{shahabuddin2019mimo}. However, their performance starts to become unstable when this ratio gets smaller. Therefore, the focus has been again shifting towards exact inversion-based linear detectors for their guaranteed performance and robustness~\cite{Prabhu_ASIC,shahabuddin2018programmable,shahabuddin2014adaptive,suikkanen2013study}. As the semiconductor technology has also matured greatly over the past decade, the focus has been shifting towards applying better system design than saving every possible logic gate. Telecommunication industries are not always interested in saving logic area by adopting an unstable and risky solutions for their products. Therefore, we envision that exact inversion-based solutions will be more popular in the future for massive MIMO VLSI community.

Matrix decomposition algorithms have been extensively utilized for the matrix inversion procedure of small-scale MIMO detection. They provide better numerical stability over straightforward inversion methods. In addition, they help to achieve a modular design, where the whole procedure can be divided and distributed between different developers. A few popular matrix decomposition algorithms for linear detection are QR, Cholesky and LDL decomposition. Despite the importance of the matrix decomposition algorithms, their complexity analysis for massive MIMO and comparison with existing AID is lacking in the literature. An analysis of matrix decomposition algorithms for small-scale MIMO detection can be found in~\cite{Sadiq_paper}. A comparison of explicit vs. implicit approximate inversion has been provided in~\cite{wu2018implicit}. However, \cite{Sadiq_paper} does not address the complexity from massive MIMO perspective and \cite{wu2018implicit} does not focus on the complexity of matrix decomposition algorithms. 

In this paper, we address this timely topic and analyze the computation complexity of matrix decomposition algorithms for massive MIMO. We also compare them with the state-of-art AID for the massive MIMO. Several hard output simulations to support our premise regarding different detection mechanisms are also presented in this paper. We believe these results will provide an important guideline to the VLSI designers to select the appropriate algorithms that is suitable to the product requirements. 

The rest of the paper is organized in the following way: In Section II, we discuss the system model of a massive MIMO system. We also discuss linear detection, non-linear detection in this section. In Section III, we introduce AID and provide simulation results to demonstrate their instability for different antenna configurations. In Section IV, we present different matrix decomposition algorithms, and their application in different types of detection methods with simulation results. In Section V, we present the complexity analysis and compare with state-of-the-art AID methods. The conclusion is drawn in Section VI. 

\section{System Model and Detection Methods}

A massive multi-user MIMO base-station (BS) with $N$ antennas is considered. We assume $U$ single antenna users are transmitting towards the BS, where $U\leq N$. Assuming that the channel between the users and BS antennas is frequency flat, the relationship between the transmit and receive vector can be characterized as 
\begin{equation}\label{gf}
    \mathbf{y}= \mathbf{H} \mathbf{x}+\mathbf{n},
\end{equation}
where $\mathbf{y} \in \mathbb{C}^{B}$ is a received signal vector, $\mathbf{x} \in \mathbb{C}^{U}$ is a transmit symbol vector, $\mathbf{H} \in\mathbb{C}^{B \times U}$ is a channel matrix, and $\mathbf{n} \in \mathbb{C}^{B}$ is a circularly symmetric complex white Gaussian noise vector with zero mean and $\sigma^2$ noise variance. The system is presented in Fig.~\ref{fig:smodel}.

\begin{figure}[h]
\centering
\includegraphics[keepaspectratio,width=.8\columnwidth]{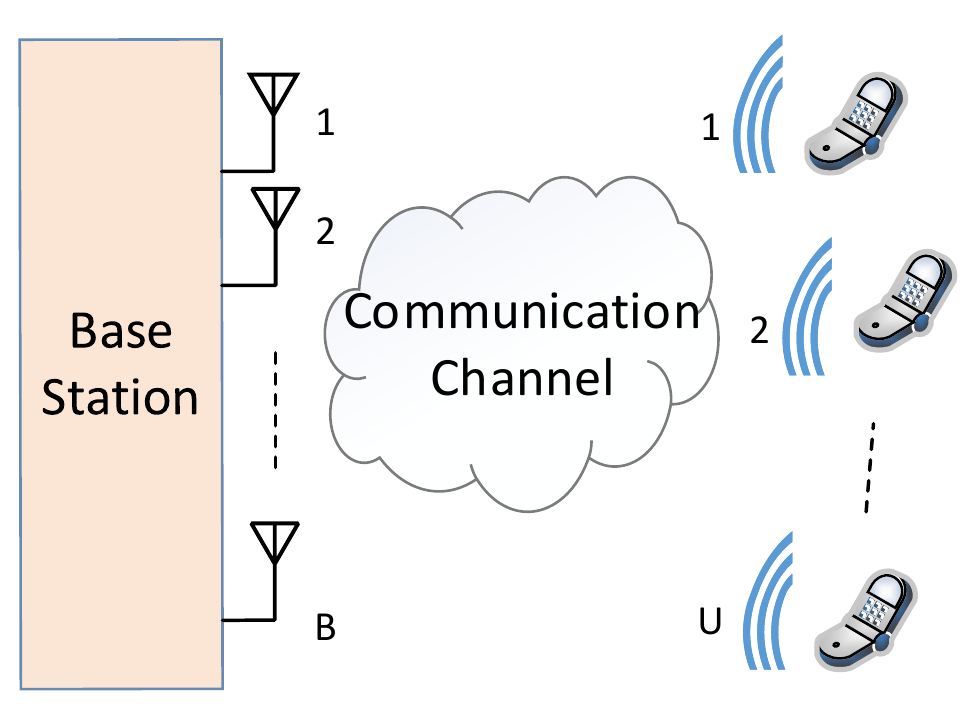}
\caption{Massive MIMO system model.}
\label{fig:smodel}
\end{figure}

\subsection{Linear Detection}

A MIMO symbol detector determines the transmitted symbol vector $\mathbf{x}$ from the received signal vector $\mathbf{y}$. The two most popular linear detection are zero-forcing (ZF) and linear minimum mean-square error (MMSE) equalization. The ZF algorithm does not consider the effect of noise vector $\mathbf{n}$, inverts the channel matrix $\mathbf{H}$ to determine the transmitted vector. Therefore, the ZF detection can be expressed as
\begin{equation}
\mathbf{\tilde{x}}_{\text{ZF}}= \mathbf{H}^{\dagger}\mathbf{y}=(\mathbf{H}^\mathbf{H}\mathbf{H})^{-1}\mathbf{H}^\mathbf{H}\mathbf{y},
\end{equation}
where $\mathbf{H}^{\dagger}$ is a pseudo-inverse of $\mathbf{H}$. The ZF detector requires an inversion of the \textit{Gramian} matrix, $\mathbf{G}$, where $\mathbf{G}_{\text{ZF}}=\mathbf{H}^\mathbf{H}\mathbf{H}$. 

The MMSE detector is an improvement over the ZF, which takes the noise into account. MMSE detection can be expressed as
\begin{equation}
\mathbf{\tilde{x}}_{\text{MMSE}}=(\mathbf{H}^\mathbf{H}\mathbf{H}+\sigma^2\mathbf{I}_U)^{-1}\mathbf{H}^\mathbf{H}\mathbf{y},
\end{equation}
where $\mathbf{I}_U$ is the $U \times U$ identity matrix. The \textit{Gramian} matrix is modified with a regularization by noise variance for MMSE, i.e., $\mathbf{G}_{\text{MMSE}}=\mathbf{H}^\mathbf{H}\mathbf{H}+\sigma^2\mathbf{I}_U$.

\subsection{Non-linear Detection}
The earliest massive MIMO detectors are iterative non-linear detectors, which are initialized with a linear detector, and in subsequent iterations tries to update the results. Example of such detectors include likelihood ascent search (LAS), reactive tabu search (RTS), CHEMP etc. We present a non-linear detector called ADMM-based Infinity-Norm Detection (ADMIN), proposed in~\cite{ADMMIscas}, to show how these non-linear detectors for massive MIMO works in general.

ADMIN solves a box-constrained detection problem with a convex optimization method called alternating direction method of multipliers (ADMM). The ADMM method can be used to solve a convex problem by breaking it into smaller problems in an iterative fashion. The first iteration of ADMIN has been presented as,
\begin{equation}\label{admin}
\mathbf{\tilde{x}}_{\text{ADMIN}}=(\mathbf{H}^H\mathbf{H} + \beta \mathbf{I})^{-1}(\mathbf{H}^H\mathbf{y} + \beta (\mathbf{z} - \pmb\lambda)),
\end{equation}
where $\beta$ is an scaled form of $\sigma^2\mathbf{I}$. The values of $\mathbf{z}$ and $\pmb\lambda$ vectors are calculated with two other iterations of ADMIN. When $\mathbf{z}$ and $\pmb\lambda$ are initialized with zeros, (\ref{admin}) resembles the MMSE equation. Thus, the ADMIN detection method is also based on an inversion of a regularized \textit{Gramian} matrix, $\mathbf{G}_{\text{ADMIN}}=\mathbf{H}^H\mathbf{H} + \beta \mathbf{I}$.

\section{Instability of Approximate Inversion-based Detection}

The exact inversion of the \textit{Gramian} can be complex when the number of users increases. For example, \textit{Gramian} of a 16-user system ($U=16$) will be a $16\times16$ matrix. Several AID became popular in the VLSI community over the past decade due to their less complexity. We present three such detectors in this section and analyze their error-rate performance with Matlab simulations.

\subsection{Neumann Series Approximation}

\textit{Gramian} matrix $\textbf{G}$ can be decomposed into a diagonal matrix (\textbf{X}) and off-diagonal matrix (\textbf{E}) as $\textbf{G} = \textbf{X} + \textbf{E}$. The Neuman series approximation (NSA)~\cite{WYWDCS2014} of such a system can be expressed as
\begin{equation}
\textbf{G}^{-1} = \sum_{t = 0}^{\infty }\left ( -\textbf{X}^{-1}\textbf{E} \right )^{t}\textbf{X}^{-1}.
\label{Eq. 16}
\end{equation}
NSA only use inversions of the diagonal matrix, $X$, which can be achieved with reciprocals of the diagonal elements. Both precision and complexity of the inversion increases with more iterations. 

\subsection{Gauss-Seidel Method}

In Gauss-Seidel (GS) method, $\textbf{G}$ can be decomposed as $\textbf{G} = \textbf{D} + \textbf{L} + \textbf{R},$ where \textbf{D}, \textbf{L} and \textbf{R} are the diagonal component, the strictly lower triangular component, and strictly upper triangular component, respectively. The GS can be used to estimate the transmitted signal vector $\hat{\textbf{x}}$ as
\begin{equation}
\hat{\textbf{x}}^{\left ( t \right )} = \left ( \textbf{D} + \textbf{L} \right )^{-1}\left ( \hat{\textbf{x}}_{\text{MF}} - \textbf{R}\hat{\textbf{x}}^{\left ( t-1 \right )} \right ),
\label{Eq. 22}
\end{equation}
where $\hat{\textbf{x}}_{\text{MF}}=\mathbf{H}^H\mathbf{y}$ is a matched filter~\cite{wu2016efficient}. 

\subsection{Conjugate Gradient Method}
Conjugate Gradient (CG) is another approximate method used in the MIMO detection which can be formulated as
\begin{equation}
\hat{\textbf{x}}^{(t+1)} = \hat{\textbf{x}}^{(t)} + \alpha ^{(t)}\textbf{p}^{(t)},
\label{Eq. 27}
\end{equation}
where $\textbf{p}^{(n)}$ is the conjugate direction with respect to the Gramian matrix and $\alpha ^{(n)}$ is a scalar parameter which is commonly known as the step size~\cite{CGChina}.

\subsection{Error-rate Performance}
We present error-rate simulation results for exact inversion-based MMSE, and approximate inversion-based NSA, GS, and CG detection methods. We have considered $t=3$ iterations of NSA, GS and CG detection methods. The bit error rate (BER) of the detectors with respect to signal-to-noise (SNR) ratios is shown in Figs 2-4. Here, we used 10,000 Monte-Carlo trials for all simulations. The modulation scheme for these simulations is 64QAM. We consider an i.i.d. Rayleigh fading channel between the BS and users. In Fig. 2, the BER for 256-antenna BS and 16 users is presented. Here, the NSA, GS, and CG present very similar performance and perform as well as the exact inversion-based MMSE. In Fig. 3, simulations have been taken for 32 antenna BS and for 16 users. Here, all AID performs poorly and can not detect received symbol vectors properly. In Fig. 4, simulations have been taken for 64 BS antennas and 16 users. Here, NSA and CG detectors can not detect received symbol. However, GS detection method provides good performance in this scenario where the MMSE provides only 2~dB gain over the GS in this scenario. 

\begin{figure}[h]
\centering
\includegraphics[keepaspectratio,width=.8\columnwidth]{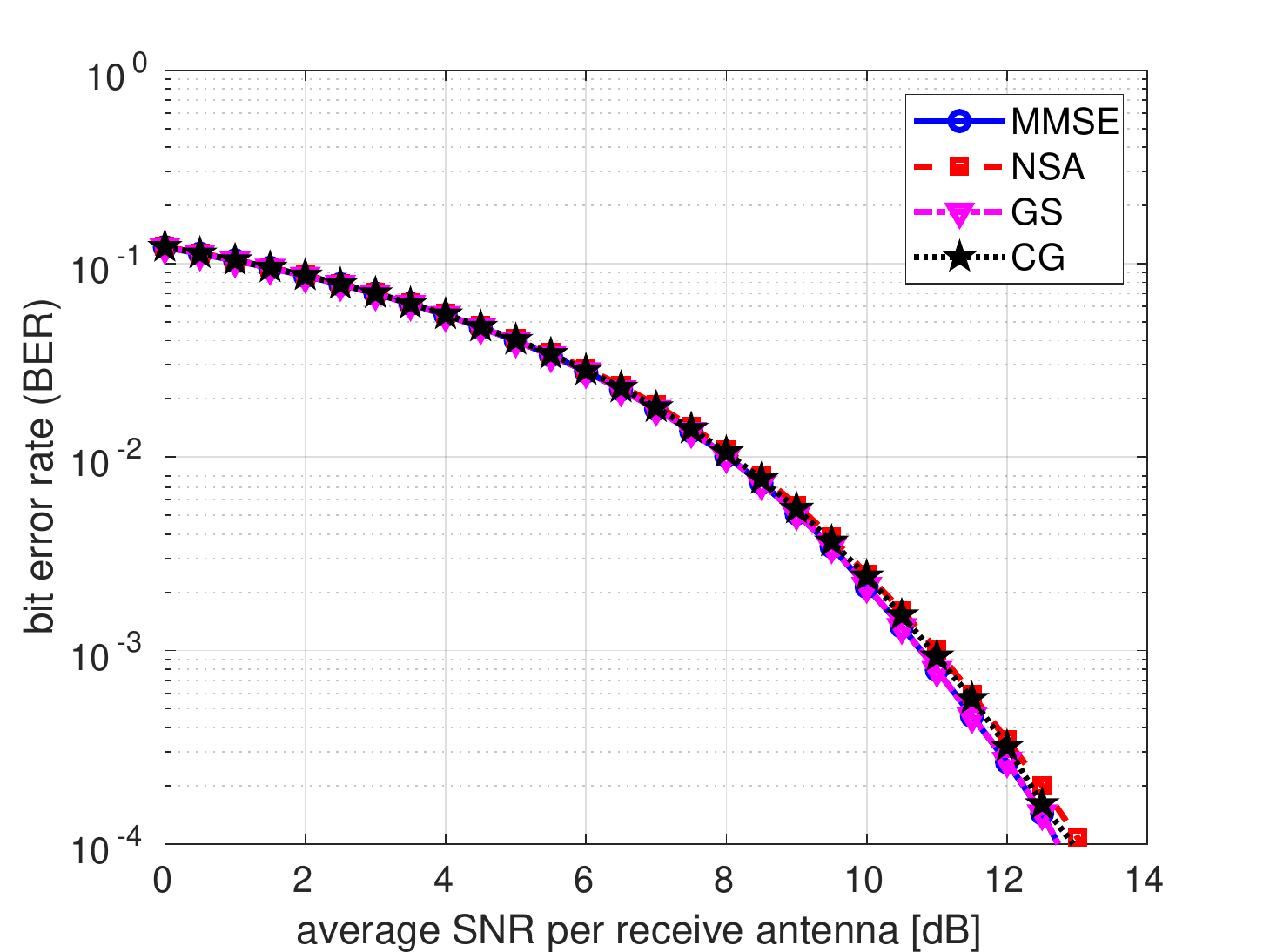}
\caption{Detector performance for 256 BS antennas and 16 users with 64-QAM.}
\label{fig:smodel}
\end{figure}

\begin{figure}[h]
\centering
\includegraphics[keepaspectratio,width=.8\columnwidth]{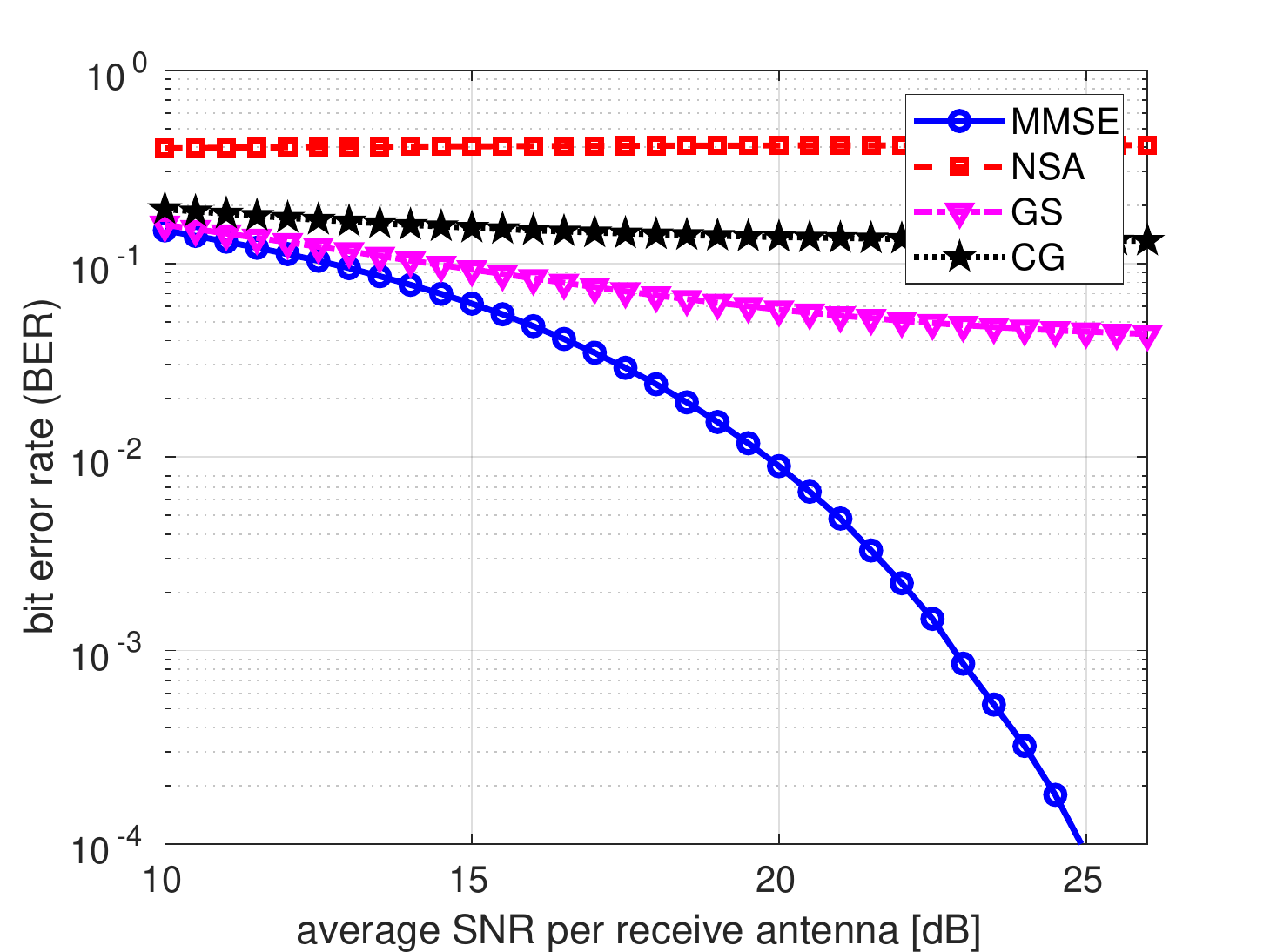}
\caption{Detector performance for 32 BS antennas and 16 users with 64-QAM.}
\label{fig:smodel}
\end{figure}

\begin{figure}[h]
\centering
\includegraphics[keepaspectratio,width=.8\columnwidth]{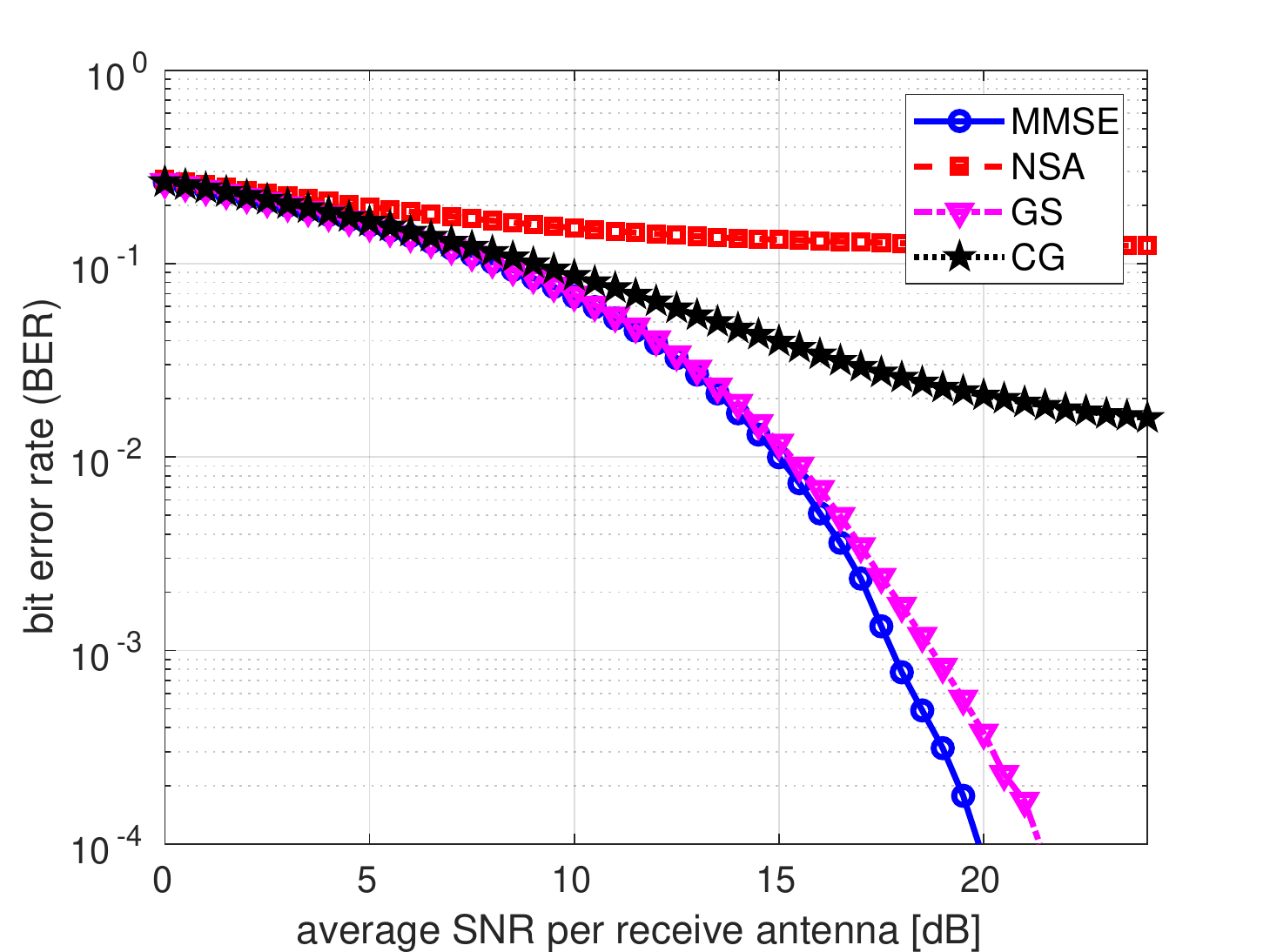}
\caption{Detector performance for 64 BS antennas and 16 users with 64-QAM.}
\label{fig:smodel}
\end{figure}

It is evident from the simulations above that all AID works very well and can provide a competitive solution compared to MMSE when the ratio between the BS antennas and number of users is high. Therefore, for such a system, it is possible to select low complexity AID methods. However, RF chains and front-end logic associated with the antennas are expensive. Therefore, having large number of antennas for a very small number of users are not always feasible from design and cost perspective. When the ratio of numbers between BS antennas and users becomes smaller, performance of AID starts to deteriorate as seen in Figs. 3 and 4. In extreme cases, when the ratio becomes close or equal to 1, these detectors do not function at all and fail to detect incoming symbol vectors. In a nutshell, the performance of these detectors is not stable and as the industry has to support various configurations for their product, such detectors might not be the ideal solution for them. Therefore, exact inversion-based detectors are still going to be an attractive solution for industries for years to come.   

\section{Matrix Decomposition Algorithms}
Instead of doing a straightforward matrix inversion, several matrix decomposition methods can be used which are more numerically stable. They also bring a modular design, where an inversion process can be split into parts. Here, we mainly focus on inverting the regularized \textit{Gramian} matrix, which is used in the MMSE equalization. In this section, we analyze how three popular matrix decomposition algorithms, i.e. QR, Cholesky and LDLT can be applied to find the MMSE solution. 

\subsection{MMSE with QR}

QR decomposition algorithm decomposes a $U\times U$ matrix $\mathbf{A}$ as $\mathbf{A}=\mathbf{QR}$. Here, $\mathbf{Q}$ is an $U\times U$ unitary matrix with orthogonal columns and $\mathbf{R}$ is an $U\times U$ upper triangular matrix with non-zero diagonal elements. QR decomposition can be applied to obtain the MMSE solution as
\begin{align}\nonumber\label{eqqr1}
&\mathbf{\tilde{x}}_{\text{MMSE}}=(\mathbf{G}_{\text{MMSE}})^{-1}\mathbf{H}^\mathbf{H}\mathbf{y}\\
&=(\mathbf{Q}\mathbf{R})^{-1}\mathbf{H}^\mathbf{H}\mathbf{y}\nonumber
=\mathbf{R}^{-1}\mathbf{Q}^\mathbf{H}\mathbf{H}^\mathbf{H}\mathbf{y}.\nonumber
\end{align}
There are several algorithms to compute the QR decomposition, such as, Gram-Schmidt process, Householder transformations, Givens rotation etc. We focus on Gram-Schmidt process in this paper.

\subsection{MMSE with Cholesky}

Cholesky is another popular decomposition algorithm which utilizes a $U\times U$ lower triangular matrix $\mathbf{L}$ to decompose a $U\times U$ matrix $\mathbf{A}$ as $\mathbf{A}=\mathbf{L}\mathbf{L}^\mathbf{H}$. The MMSE can be calculated as
\begin{align}\nonumber
&\mathbf{\tilde{x}}_{\text{MMSE}}=(\mathbf{G}_{\text{MMSE}})^{-1}\mathbf{H}^\mathbf{H}\mathbf{y}=(\mathbf{L}\mathbf{L}^\mathbf{H})^{-1}\mathbf{H}^\mathbf{H}\mathbf{y}\nonumber\\
&\Longrightarrow \mathbf{L}\mathbf{L}^\mathbf{H}\mathbf{\tilde{x}}_{\text{MMSE}} = \mathbf{H}^\mathbf{H}\mathbf{y}\nonumber\\
&\Longrightarrow \mathbf{L}\mathbf{\tilde{z}} = \mathbf{H}^\mathbf{H}\mathbf{y} \quad \text{and} \quad\mathbf{L}^\mathbf{H}\mathbf{\tilde{x}}_{\text{MMSE}}=\mathbf{\tilde{z}}.\nonumber
\end{align}
Here, $\mathbf{\tilde{z}}$ can be calculated by forward substitution and $\mathbf{\tilde{x}}$ can be solved by backward substitution. 

\subsection{MMSE with LDL}
LDL is another popular decomposition algorithm which decomposes a $U\times U$ matrix $\mathbf{A}$ as $\mathbf{A}=\mathbf{L}\mathbf{D}\mathbf{L}^\mathbf{H}$, where $\mathbf{D}$ is a diagonal matrix and $\textbf{L}$ is a lower triangular matrix with zeros as diagonal. The MMSE can be calculated with the LDL as
\begin{align}
&\mathbf{\tilde{x}}_{\text{MMSE}}=(\mathbf{G}_{\text{MMSE}})^{-1}\mathbf{H}^\mathbf{H}\mathbf{y}=(\mathbf{L}\mathbf{D}\mathbf{L}^\mathbf{H})^{-1}\mathbf{H}^\mathbf{H}\mathbf{y}\nonumber\\
&\Longrightarrow \mathbf{L}\mathbf{D}\mathbf{L}^\mathbf{H}\mathbf{\tilde{x}}_{\text{MMSE}} = \mathbf{H}^\mathbf{H}\mathbf{y}\nonumber\\
&\Longrightarrow \mathbf{L}\mathbf{\tilde{z}} = \mathbf{H}^\mathbf{H}\mathbf{y} \quad \text{and} \quad\mathbf{L}^\mathbf{H}\mathbf{\tilde{x}}_{\text{MMSE}}=\mathbf{D}^{-1}\mathbf{\tilde{z}}.\nonumber
\end{align}
Similar to Cholesky, $\mathbf{\tilde{z}}$ can be calculated by forward substitution and $\mathbf{\tilde{x}}$ can be solved by backward substitution. 

\subsection{Error-rate Performance}
We present error-rate performance of MMSE, ADMIN and single-input multiple-output (SIMO) in Fig. 5 and 6. Similar to earlier simulations, 10,000 Monte-Carlo trials are used. The communication channel between BS and users is assumed to be i.i.d. Rayleigh Fading channel.
The MMSE utilizes QR and ADMIN utilizes LDL decomposition in these simulations. We take $t=5$ iterations of ADMIN in this simulations. In Fig. 5, the detectors are simulated for 32 BS antennas supporting 32 users transmitting with 64-QAM. It can be noticed that even the exact matrix inversion-based MMSE can not provide a satisfactory performance with high power, i.e., can not bring the error-rate below $10^{-2}$ with approximately 35~dB of SNR. ADMIN provides about 5~dB gain over MMSE, which is a significant margin for communication systems using higher order modulation. Fig. 4 shows, when the ratio of numbers between BS antennas and users is small, non-linear detectors are required for necessary performance gain. Most of these non-linear detectors are based on exact inversion and matrix decomposition algorithms will be crucial for their implementation. The non-linear detectors can also provide very high gain compared to MMSE for other scenarios. In Fig. 5, ADMIN with $t=5$ iterations can provide around 10~dB gain over MMSE algorithm. In this case, the difference between the SIMO bound and ADMIN is less than 10~dB.

\begin{figure}[h]
\centering
\includegraphics[keepaspectratio,width=.8\columnwidth]{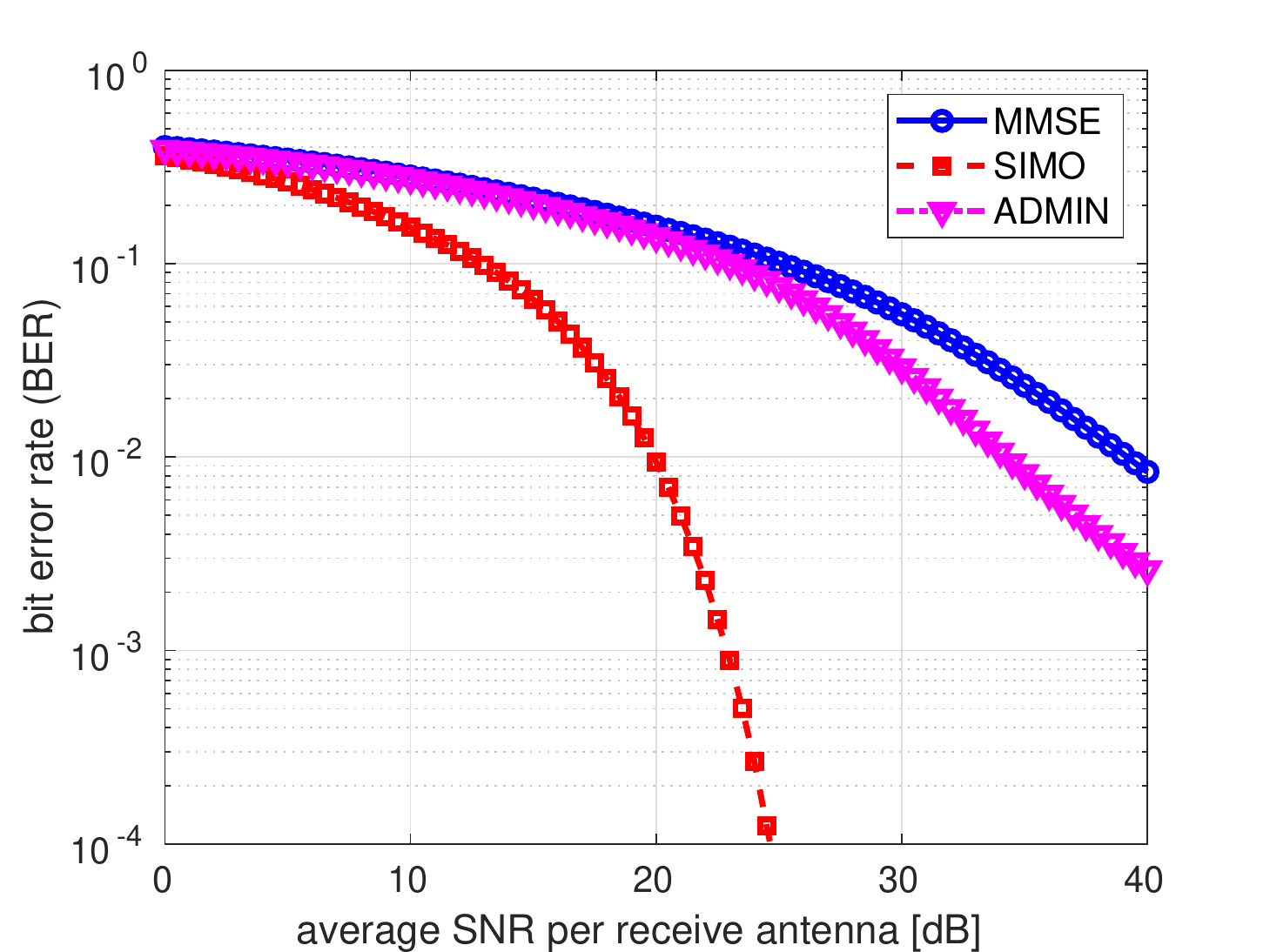}
\caption{Detector performance utilizing matrix decomposition for 32 BS antennas and 32 users with 64-QAM.}
\label{fig:smodel}
\end{figure}

\begin{figure}[h]
\centering
\includegraphics[keepaspectratio,width=.8\columnwidth]{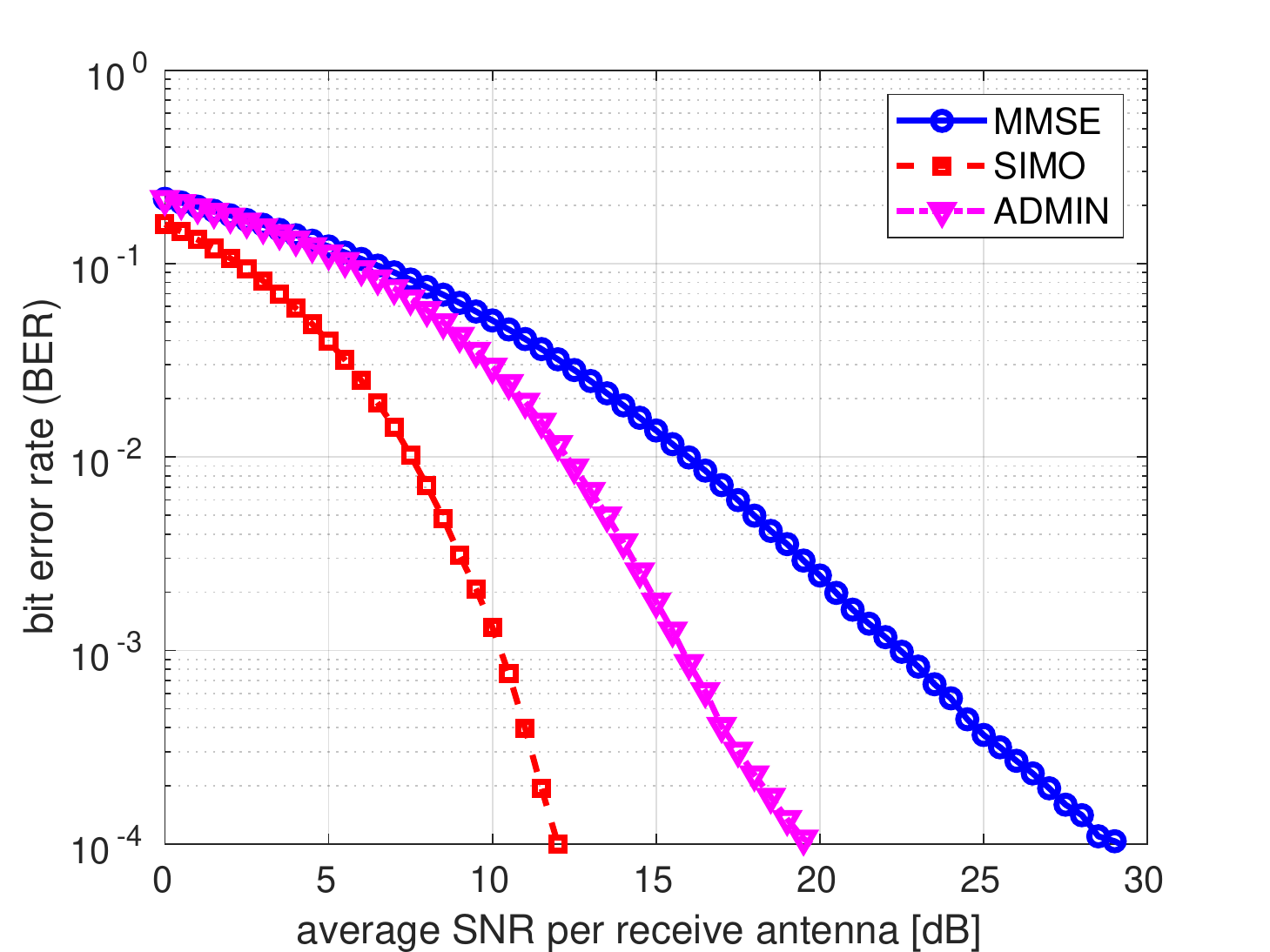}
\caption{Detector performance utilizing matrix decomposition for 32 BS antennas and 32 users with QPSK.}
\label{fig:smodel}
\end{figure}

\section{Complexity Analysis}

We analyze the complexity of matrix decomposition algorithms in this section. First, we explain in detail how we computed the complexity of QR decomposition based on classical Gram-Schmidt process. We calculate the complexity of Gram-Schmidt applied on a $U\times U$ matrix because the Gramian is always square in the MMSE. Real multiplications dictate the overall complexity of an algorithm as they take significantly more logic than additions or subtractions. Therefore, we focus on real multiplications for comparing the algorithms. Classical Gram-Schmidt algorithm is presented in Algorithm~\ref{algo1}.

\begin{algorithm}[t]
	\caption{Classical Gram-Schmidt}
	\label{algo1}
	\begin{algorithmic}
		\STATE{\textbf{\textit{input}}: $\mathbf{A}$}
	    \STATE{\textbf{\textit{outputs}}: $\mathbf{Q}$, $\mathbf{R}$  }
		\STATE{1: $\mathbf{Q}=\mathbf{A}$}
		\STATE{2: $\textbf{for}$ $i = 1,\ldots,U $}
		\STATE{3: \hspace{4 mm} $r_{i,i} = \|\mathbf{q}_i\|^2$}
		\STATE{4: \hspace{4 mm} $\mathbf{q}_i=\mathbf{q}_i/r_{i,i}$ }
		\STATE{5: \hspace{4 mm} $\textbf{for}$ $j = i+1,\ldots,U $}
		\STATE{6: \hspace{8 mm} $r_{i,j}=\mathbf{q}_i^{H}\mathbf{q}_j$    }
        \STATE{7: \hspace{8 mm} $\mathbf{q}_j=\mathbf{q}_j-r_{i,j}\mathbf{q}_i$    }
		\STATE{8: \hspace{4 mm} \textbf{end}  }
		\STATE{9: \textbf{end}  }
	\end{algorithmic}
\end{algorithm}

In line 3 of Algorithm~\ref{algo1}, norm of a vector is computed as $r_{i,i} = \|\mathbf{q}_i\|^2$. For norm calculation, we need a total of $U^2$ number of complex multiplications. Here, $U$ number of multiplications are required for a single norm and we need a total of $U$ norms. For $\mathbf{q}_i=\mathbf{q}_i/r_{i,i}$, a total of $U$ reciprocals needed to compute $\frac{1}{r_{i,i}}$ and these real-valued reciprocals will be multiplied $U^2$ times to the complex valued $\mathbf{q}_i$. Thus, a total $2U^2$ real multiplications are needed for line 4. Next, $r_{i,j}=\mathbf{q}_i^{H}\mathbf{q}_j$ implies a vector–vector multiplication where each vector has $U$ complex elements. Therefore, we need $U$ number of complex multiplications for a single $\mathbf{q}_j$. Here, we have $\frac{U(U-1)}{2}$ number of off-diagonal elements in a $U\times U$ matrix. So, we have $\left(\frac{U(U-1)}{2}\right)U$ number of complex multiplications for $r_{i,j}\mathbf{q}_i$. Similarly, we get a total of $\left(\frac{U(U-1)}{2}\right)U$ multiplication for $r_{i,j}\mathbf{q}_i$. The operations are presented in Table~\ref{tab1}.
\begin{table}[h]
\centering

\caption{Operations of classical Gram-Schmidt}
\label{tab1}
{\renewcommand{\arraystretch}{1.6}
\begin{tabular}{|l|l|l|}
\hline
Line & Equation & Operations\\
\hline
3 & $r_{i,i} = \|\mathbf{q}_i\|^2$ & $U^2$ complex multiplication \\
\hline
4 & $\mathbf{q}_i=\mathbf{q}_i/r_{i,i}$ & $2U^2$ real multiplication \\
\hline
6 &  $r_{i,j}=\mathbf{q}_i^{H}\mathbf{q}_j$ & $\left(\frac{U(U-1)}{2}\right)U$ complex multiplication\\
\hline
7 &  $\mathbf{q}_j=\mathbf{q}_j-r_{i,j}\mathbf{q}_i$ & $\left(\frac{U(U-1)}{2}\right)U$ complex multiplication\\
\hline
\end{tabular}}
\end{table}

Therefore, we get the total number of complex multiplication (CM) for Gram-Schmidt as
\begin{align}
    \text{CM}&= U^2  +  \frac{(U(U-1))}{2}U+\frac{(U(U-1))}{2}U\nonumber\\
    &=U^2  +2\frac{U^2(U-1)}{2}\nonumber \\
    &=U^2  + U^3 - U^2 = U^3\nonumber
\end{align}
We assume, a complex-complex scalar multiplication hardware unit utilizes four real multiplication units and a real-complex scalar multiplication unit utilize two real multiplication units. Therefore, we get $4U^3$ real multiplication from the $U^3$ complex multiplication. We add this with number of real multiplications (RM) for line 4, which leads to
\begin{equation}\nonumber
\text{RM} = 4U^3 + 2U^2 = U^2(4U+2).
\label{Eq. rm}
\end{equation}

\begin{algorithm}[t]
	\caption{Cholesky Decomposition}
	\label{algo2}
	\begin{algorithmic}
		\STATE{\textbf{\textit{input}}: $\mathbf{A}$}
	    \STATE{\textbf{\textit{outputs}}: $\mathbf{L}$ }
		\STATE{1: $\textbf{for}$ $i = 1,\ldots,U $}
		\STATE{2: \hspace{4 mm} $j = 1,\ldots,i-1$}
		\STATE{3: \hspace{4 mm} $ \mathbf{L}_{i,i}= \sqrt{\mathbf{A}_{i,i}-\mathbf{L}_{i,j}\mathbf{L}_{i,j}^*}$}
		\STATE{4: \hspace{4 mm} $\textbf{for}$ $k = i+1,\ldots,U $}
		\STATE{5: \hspace{8 mm} $\mathbf{L}_{k,i}=\cfrac{1}{\mathbf{L}_{i,i}^*}\left(\mathbf{A}_{k,i}-\mathbf{L}_{k,j}\mathbf{L}_{i,j}^*\right)$    }
		\STATE{6: \hspace{4 mm} \textbf{end}  }
		\STATE{7: \textbf{end}  }
	\end{algorithmic}
\end{algorithm}

Similarly, we calculate the number of multiplications required for Cholesky decomposition from Algorithm~\ref{algo2}. A total of $\frac{U(U-1)}{2}$ complex multiplication is required to compute the diagonal elements of $\mathbf{L}$ (line 3). The off-diagonal elements require a total of $\frac{U^3-3U^2+2U}{6}$ complex multiplication (line 5). Therefore, the total number of complex multiplication is
\begin{equation}\nonumber
        \text{CM}= \frac{U(U-1)}{2}+\frac{U^3-3U^2+2U}{6} = \frac{U^3-U}{6}
\end{equation}
Computation of Cholesky also requires a $U$ number of reciprocals of $\cfrac{1}{\mathbf{L}_{i,i}^*}$, which are then multiplied in line 5. Therefore, we require an additional $U(U-1)$ real multiplications. Therefore, the total number of real-multiplication is
\begin{equation}\nonumber
        \text{RM}= 4\frac{U^3-U}{6}+ U(U-1) = \frac{2U^3+3U^2-5U}{3}.
\end{equation}
The complexity of LDL decomposition is very similar to that of Cholesky with an additional $4U(U-1)$ real multiplications. We provide Table~\ref{tab1b} with more detail of the operations required for the matrix decomposition algorithms. The number of real multiplication, addition and subtractions for QR is significantly higher than Cholesky and LDL decomposition. The complexity of Cholesky and LDL are almost similar except the additional multiplication required for LDL and the square root operations required for Cholesky.

\begin{table*}[t]
	\centering
	\caption{Operations in detail for matrix decomposition algorithms}
	\label{tab1b}
	\begin{tabular}{|c|c|c|c|c|c|c|}
		\hline
		No. of users& Algorithm& Square root & Reciprocal & Multiplication & Addition & Subtraction \\
		\hline
		\multirow{3}{*}{8} 
		&Gram-Schmidt &  8 & 8 & 2432 & 604 & 576  \\
		&Cholesky     &  8 & 8 & 392  &   42 &  36  \\
		&LDL          & -  & 8 & 560 &  42 &  36  \\
		\hline
		\multirow{3}{*}{16} 
		&Gram-Schmidt &  16 & 16 & 17920 & 4532 & 4352  \\
		&Cholesky     &  16 & 16 & 2960  & 210  & 136  \\
		&LDL          &   - & 16 & 3680 & 210  & 136 \\
		\hline
		\multirow{3}{*}{32} 
		&Gram-Schmidt & 32  & 32   & 137216 & 34660 & 33792 \\
		&Cholesky     & 32 & 32 & 22816 & 930 & 528  \\
		&LDL          & - & 32 & 25792 & 930 & 528 \\
		\hline
	\end{tabular}
\end{table*}

\subsection{Comparison against approximate inversion-based detectors}
A comparison of the matrix decomposition algorithms and the AID methods is presented in Table~\ref{tab_complexity}. All the matrix decomposition algorithms and NSA scale with $U^3$ while GS and CG methods scale with $U^2$ in terms of operations. The cubic complexity of NSA originates from the matrix multiplications associated with every iteration $t$ in $(\textbf{X}^{-1}\textbf{E})^{t}$. For example, for $t=3$, we have to compute a $\mathbf{P}*\mathbf{P}*\mathbf{P}$ where $\mathbf{P}=\textbf{X}^{-1}\textbf{E}$. In Fig. 7, matrix decomposition algorithms and the AID methods are compared in terms of numbers of real multiplication required for each algorithm. Here, the AIDs applied $t=3$ iterations each. From this figure, the QR decomposition and the NSA are significantly more complex than the other algorithms. Cholesky and LDL provide similar complexity, which is also supported by the operation numbers of Table~\ref{tab1b}. GS and CG methods require the least number of multiplications for their execution.

\begin{figure}[h]
\centering
\includegraphics[keepaspectratio,width=1\columnwidth]{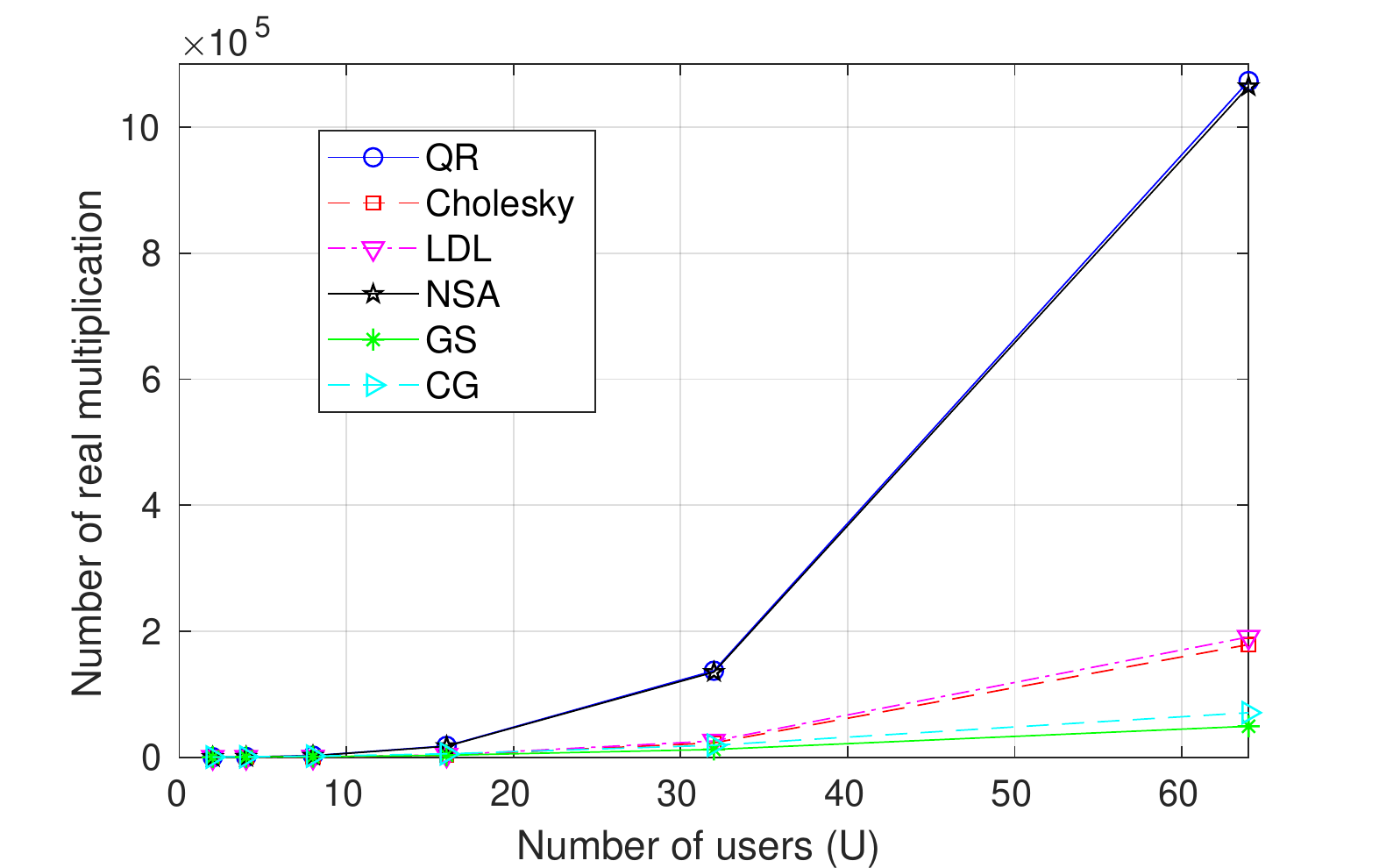}
\caption{Complexity comparison of matrix decomposition algorithms and approximate inversion-based detectors.}
\label{fig:smodel}
\end{figure}

\begin{table}[t]
	\centering
	\caption{Complexity comparison}
	\renewcommand\arraystretch{1.8}
	\label{tab_complexity}
	\begin{tabular}{|l|l|}
		\hline
		Algorithm & Computational complexity \\
		\hline
		QR &  $U^2(4U+2)$  \\
		\hline
		Cholesky &
		$\frac{1}{3}(2U^3+3U^2-5U)$  \\
		\hline
		LDL &
		$\frac{1}{3}(2U^3+12U^2-14U)$  \\
		\hline
		NSA&
		$(t-1)(2U^3+2U^2-2U)$  \\
		\hline
		GS &
		$6tU^2$  \\
		\hline
		CG & $(t+1)(4U^2+20U$)  \\
\hline	
\end{tabular}
\end{table}

As Cholesky and LDL provide significantly less complexity over the QR decomposition, they are ideal choice for implementation platforms. These algorithms even provide lower complexity than NSA. The error-rate performance of the CG method is not satisfactory according to Figs. 3 and 4. On the other hand, GS method is very promising as it outperforms other AIDs in terms of error rate while using low number of operations. From Fig. 7, the number of operations required for GS is less than half of Cholesky and LDL. It should be noted that the channel model utilized in this work is i.i.d. and therefore, the performance of the GS method might suffer in a more realistic channel model. We plan to compare GS, Cholesky, LDL in a more realistic channel model for our future works.

\section{Conclusion}
We presented an analysis of matrix decomposition algorithms from massive MIMO context and their suitability for VLSI design of massive MIMO detectors. We compared massive MIMO detection algorithms based on approximate inversion and conventional matrix decomposition algorithms. The results will provide a guideline to select the appropriate algorithm for the research community. We concluded that Cholesky and LDL are viable solutions for massive MIMO detection. We also found that the GS can be an attractive approximate inversion-based detector which might provide a good balance between error-rate performance and complexity.


\ifCLASSOPTIONcaptionsoff
  \newpage
\fi

\bibliographystyle{IEEEtran}
\bibliography{references}

\end{document}